\documentclass[11pt,amssymb,showpacs]{iopart}
\usepackage{latexsym}
\usepackage{graphicx}
\usepackage{color}

\begin{document}

\title[Metallicity of the $SrTiO_3$ surface induced by the evaporation of alumina]{Metallicity of the $SrTiO_3$ surface induced by room temperature evaporation of alumina}
\author{J. Delahaye and T. Grenet}
\address{Institut N\'eel, CNRS $\&$ Universit\'e Joseph Fourier, BP 166, 38042 Grenoble, France}
\ead{julien.delahaye@grenoble.cnrs.fr}
\date{\today} 

\begin{abstract}

It is shown that a metallic state can be induced on the surface of $SrTiO_3$ crystals by the electron beam evaporation of oxygen deficient alumina or insulating granular aluminium. No special preparation nor heating of the $SrTiO_3$ surface is needed. Final metallic or insulating states can be obtained depending on the oxygen pressure during the evaporation process. Photoconductivity and electrical field effect are also demonstrated.

\end{abstract}
 \bigskip
\maketitle

\section{Introduction}

$SrTiO_3$ crystals are insulators with a band gap of $3.2eV$. The ability to change them into metals is a crucial issue towards future electronic applications. Different techniques have been used through the last 50 years, such as classical doping techniques \cite{FrederiksePR64,SpinelliPRB10}, heating to high temperature in vacuum \cite{FrederiksePR64,SpinelliPRB10} or ion-bombardment of the surface \cite{ReagorNatMat05,GrossJAP11}. In these last two cases, oxygen vacancies are introduced in the crystals and release electrons for the conduction.

But the most spectacular results were observed in $LaAlO_3/SrTiO_3$ heterostructures grown by pulsed laser deposition (PLD) at high temperature (typically $800^{\circ}C$)\cite{OhtomoNature04,ThielScience06,HuijbenNatMat06}. Under specific growth conditions, a high mobility electron gas confined to the interface was found on $(100)$ oriented and $TiO_2$ terminated $SrTiO_3$ surface \cite{ThielScience06,HuijbenNatMat06,BastelicNatMat08}. The respective roles of oxygen vacancies \cite{KalabukhovPRB07,SiemonsPRL07,HerranzPRL07}, ionic interdiffusion \cite{NakagawaNatMat06,WillmottPRL07} and electronic reconstruction \cite{OhtomoNature04,ThielScience06} in the formation of this electron gas are not yet fully clarified.
Here, we demonstrate that a metallic layer can be induced at room temperature on the $SrTiO_3$ surface by using classical electron beam evaporation techniques. After a description of our experimental results, we discuss the possible origin of this metallic state.

\section{Experimental}

  We used $SrTiO_3$ crystals from Neyco, one side polished, $(100)$ oriented and $0.5mm$ thick. The polished surface was simply cleaned by successive ultrasonic bath in trichlorethylene, acetone and alcohol before being mounted in the electron beam evaporator. $20nm$ thick $Al$ contacts are evaporated first. Then, about $5nm$ of alumina or insulating granular $Al$ are deposited between the contacts, which are partly protected by a mechanical mask. The alumina layer is obtained from the evaporation of $Al_2O_3$ crystallites at $0.5A/s$ with or without $O_2$ pressure (base pressure of the evaporator $\simeq 10^{-6}mbar$). The insulating granular $Al$ layer is made by the evaporation of pure $Al$ at $1.8\AA/s$ under an $O_2$ pressure around $2 \times 10^{-5}mbar$ (see \cite{GrenetEPJB07} and the discussion below). Finally, this $\simeq 5nm$ thick layer is covered by a protective layer of alumina $95nm$ thick and evaporated at an $O_2$ pressure of $2\times 10^{-4}mbar$. For the insulating samples, the resistances were measured with a two contacts configuration. For the metallic samples, two or four contacts configurations were used and gave similar results \cite{Contacts}. The current-voltage curves were all linear at low voltages .

\section{Results and discussion}

\subsection{Samples electrical properties}

\begin{table}[h]
\caption{\label{table1} $R_s$ values obtained at $300K$ for different evaporated materials and $O_2$ pressures during the evaporation. The "light" value was measured under an indirect daylight. The "dark" value was measured $20000s$ after a stay in a closed metallic box.}
\begin{indented}
\item[]\begin{tabular}{ccccc}
\br
Sample & Evap. mat. & $O_2$ pressure & $R_s$ "light" & $R_s$ "dark" \\
\mr
1 & $Al_2O_3$ & $<3\times 10^{-6}$ & $20k\Omega$ & $22k\Omega$ \\
2 & $Al_2O_3$ & $1\times 10^{-5}$ & $30M\Omega$ & $150M\Omega$ \\
3 & $Al_2O_3$ & $4\times 10^{-5}$ & $80M\Omega$ & $1.5G\Omega$ \\
4 & $Al_2O_3$ & $2\times 10^{-4}$ & $800M\Omega$ & $40G\Omega$ \\
5 & $Al$ & $3\times 10^{-5}$ & $10k\Omega$ &  \\
\br
\end{tabular}
\end{indented}
\end{table}

\begin{figure}[h]
    \includegraphics[width=8cm]{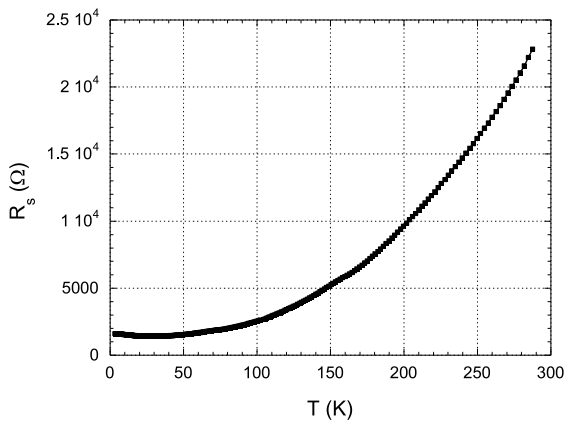}
    \caption{Temperature dependence of the sheet resistance $R_s$ for sample 1.}\label{Figure1}
    \end{figure}

Typical results for the sheet resistances $R_s$ at room $T$ are reported in Table \ref{table1}. It is seen that the resistance strongly depends on the oxygen pressure during deposition and that resistances as low as a few tens of $k\Omega$ can be obtained. Samples 1 and 5 are indeed metallic, as highlighted by the $R-T$ curve of sample 1 plotted in Fig. \ref{Figure1} (resistance ratio $R_{300K}/R_{4K}$ is about 10). For the sample 1, Hall effect measurement at $300K$ gives a negative sheet charge carrier density of $\simeq 5\times 10^{13} cm^{-2}$ and a mobility of $\simeq 6 cm^2V^{-1}s^{-1}$ \cite{Contacts}. All the other films are insulating, and their resistances were not measurable at $4K$ (larger than $10^{12}\Omega$). Moreover, they display a large and slow photoconductivity effect. For the metallic ones, the "light" contribution represents only a small fraction of the resistance, typically $10\%$. A similar photoconductivity attributed to photocarrier injection was also reported in $SrTiO_3/LaAlO_3$ heterostructures made by PLD \cite{HuijbenNatMat06,HuijbenThesis06,ThielThesis09}.

\begin{figure}[h]
    \includegraphics[width=8cm]{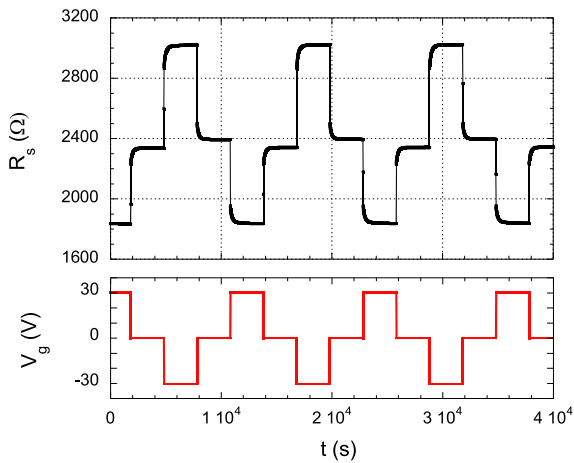}
    \caption{$V_g$ modulation of $R_s$ for sample 1 at $4.2K$. The source-drain voltage was kept smaller than $100mV$ and $V_g$ was successively and repeatedly fixed to $30V$, $0V$, $-30V$ and $0V$.}\label{Figure2}
    \end{figure}

We have tested the influence of a gate voltage $V_g$ on the resistance of the samples. The backside of the $SrTiO_3$ substrate was covered with silver paint and used as the gate, the bulk of the $SrTiO_3$ crystal being the gate insulator. Due to its large dielectric constant at low $T$ (up to $20000$ at $4K$ \cite{MullerPRB79}), sheet charge carrier densities as high as $7\times 10^{12}cm^{-2}$ can be induced by a $V_g$ value of $30V$. Typical result for the metallic sample 1 is shown at $4K$ in Fig. \ref{Figure2}. A clear modulation of the resistance is visible, with a small memory effect when the gate voltage is switched back to $0V$. At room $T$, the relative resistance modulation is smaller and very slow. For the insulating samples, no measurable resistance can be induced at low $T$ by a $V_g$ value of $30V$ whereas a large but slow relative modulation of the resistance is observed at room $T$.

\subsection{What is metallic?}

\begin{figure}[h]
    \includegraphics[width=8cm]{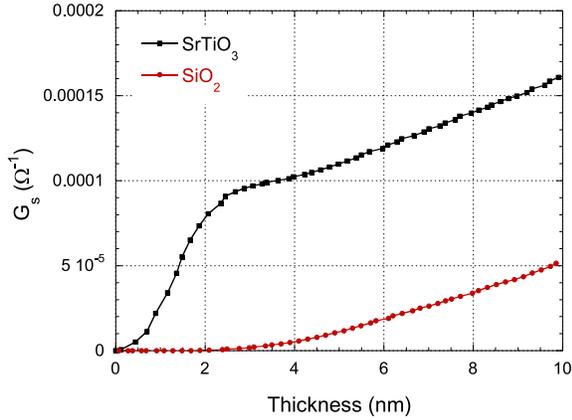}
    \caption{Sheet conductance $G_s$ measured in-situ during the deposition of a granular $Al$ layer on $SrTiO_3$ (squares) and $SiO_2$ (circles) substrates. The $Al$ rate is $1.8\AA/s$ and the $O_2$ pressure $1.8 \times 10^{-5}mbar$ for both substrates.}\label{Figure3}
    \end{figure}

The first question which arises is: what is metallic in samples 1 and 5? Is it the $SrTiO_3$ substrate or the material deposited on top? Sample 1 was obtained from the evaporation of an insulating material, alumina. Even if the alumina layer deposited on the $SrTiO_3$ substrate is most probably oxygen deficient (see below), it is still insulating. This was confirmed by evaporating alumina with the same parameters as for sample 1 but on a glass substrate: no resistance was measurable at room $T$ ($R_s > 10^{12}\Omega$). For sample 5, the response is less trivial since the granular $Al$ layer is obtained from the evaporation of a metal, $Al$, in the presence of oxygen. We know from our previous studies on sapphire or $SiO_2$ substrates that above and close to an $O_2$ pressure of $\simeq 1.5 \times 10^{-5}mbar$, the deposited films are insulating and made of nanometric $Al$ grains separated by amorphous alumina regions, a material called granular $Al$ \cite{GrenetEPJB07,PetrovJVST03}. Films $20nm$ thick evaporated on $SiO_2$ substrates with the same parameters as sample 5 have $R_s$ value of $\simeq 100M\Omega$ at room $T$. Even more informative was the direct comparison between two granular $Al$ layers evaporated with the same parameters but on $SrTiO_3$ and $SiO_2$ substrates. The conductance displayed in Fig. \ref{Figure3} was measured in-situ between two metallic contacts during the evaporation process. A striking difference is seen: on $SrTiO_3$, the conductance sharply increases during the first $2-3 nm$ of deposition (amounting to $10^{-4} \Omega ^{-1}$ at $2nm$) while on $SiO_2$ the conductance remains much smaller. This increase comes from the presence of the $SrTiO_3$ substrate becoming conducting upon deposition of granular $Al$. It appears from Fig. \ref{Figure3} that the $SrTiO_3$ contribution to the conductance is almost complete after the deposition of $3nm$, which justifies our choice of $5nm$ for the thickness of the alumina or granular $Al$ layers. The increase of the conductance seen beyond $\simeq 4nm$ is similar for both substrates and reflects the granular $Al$ contribution (this contribution is significant here due to an $O_2$ pressure smaller than for sample 5).

\subsection{Role of the oxygen pressure and origin of the metallicity}

The next question we have to answer is: why the $SrTiO_3$ surface becomes metallic? It clearly appears from Table \ref{table1} that the $O_2$ pressure plays a role in the resistance of the samples. Looking at the alumina samples, the resistance increases from $20k\Omega$ to $30M\Omega$ when the $O_2$ pressure is changed from $<3\times 10^{-6}mbar$ to $1\times 10^{-5}mbar$. But when evaporating from a pure $Al$ source, a metallic state is obtained even with an $O_2$ pressure as high as $3\times 10^{-5}mbar$ (sample 5), which suggests that it is not the $O_2$ pressure but indeed the $Al/O$ ratio that matters. $Al$ is known to react strongly with oxygen, the $Al_2O_3$ phase being the stable final state. A possible explanation to our experimental results is that during the deposition of an alumina layer at low $O_2$ pressure, part of the missing oxygen is "pumped" from the $SrTiO_3$ substrate. Oxygen vacancies are indeed known to release electrons \cite{FrederiksePR64}, in agreement with the sign of the Hall effect measurements on sample 1. This pumping oxygen scenario was strengthened by two more results. First, the highest breakdown voltages of $100nm$ thick alumina layers evaporated on sapphire were obtained for an $O_2$ pressure above $1\times 10^{-4}mbar$. This suggests that the alumina layers of samples 1 to 3 are indeed oxygen deficient, whereas that of sample 4 is close to stoichiometric conditions (this is why the parameters of sample 4 were chosen for the evaporation of the $95nm$ alumina protective layer). Second, in one sample made under similar conditions to sample 1, the $Al_2O_3$ evaporation rate was stopped during $10s$ after the deposition of $1nm$ of alumina. Instead of being metallic, this sample was strongly insulating, with a $R_s$ value similar to that of the sample 4. During $10s$, the missing oxygen was probably provided by the residual $O_2$ pressure of the chamber and not by $SrTiO_3$ substrate.

\subsection{Possible mechanism for the oxygen transfer}

It is known that oxygen vacancies are introduced in the bulk of $SrTiO_3$ crystals during an annealing in a reducing atmosphere ($\simeq 900^{\circ}C$ and above)\cite{FrederiksePR64} and during the epitaxial growth of oxides at high temperature by PLD or molecular beam epitaxy under a low oxygen pressure \cite{ShimoyamaJJAP01,UedonoJAP02,HerranzAPL09,SchneiderAPL10}. In this last case, an oxygen transfer occurs from the $SrTiO_3$ substrate to its oxide overlayer, a mechanism called "autofeeding epitaxy". But out-diffusion of oxygen at room temperature has been much rarely reported. Metal oxidation and corresponding $SrTiO_3$ reduction was indeed observed at room temperature when reactive metals are deposited on top of $SrTiO_3$ surfaces under ultra high vacuum  \cite{HillJAP89,FuJPCB05}, the results being strongly dependent on the metal element \cite{FuSSR07}. The heat of formation of the corresponding metal oxide determines if the metal oxidation is energetically favorable and the metal work function influences the onset temperature of metal oxidation. The role of the metal work function was attributed to the existence of local electric fields at the metal-$SrTiO_3$ interface that favor or prevent the out-diffusion of the negatively charged oxygen ions. Metal oxidation was indeed observed under ultra high vacuum and at room temperature for $Al$ \cite{FuJPCB05}, $Y$, $Ti$ and $Ba$ \cite{HillJAP89}, metals which combine large oxide heat of formations and low metal work functions \cite{FuSSR07,NoteAl}. For $Y$, $Ti$ and $Ba$, the metal oxidation was found to be diffusion limited, with a gradual decrease of the oxygen content in the first nanometers of the metal overlayer \cite{HillJAP89}. We believe that the metallic state observed when an insulating granular $Al$ is evaporated (sample 5) is due to the same oxygen transfer mechanism. The difference with the metal oxidation discussed in Ref. \cite{HillJAP89,FuJPCB05} is that in our case oxygen from the substrate competes with the gaseous oxygen in the evaporation chamber. The conductance measurement of Fig. 3 indicates that the $SrTiO_3$ reduction saturates after the deposition of $2-3nm$ of granular $Al$, which can be explained by the limited diffusion of oxygen in the granular $Al$ overlayer. When alumina is evaporated (samples 1 to 4), $Al$ grains probably never form. But as long as the alumina layer is oxygen deficient, a similar (but weaker) reduction of the $SrTiO_3$ substrate certainly occurs. Structural and chemical investigations are now needed in order to precise the nature of the insulating overlayers and the oxygen vacancies distribution in the $SrTiO_3$ substrate when the evaporation is completed. Whether energetic species or X-rays generated by the e-beam gun play a role in the formation of the oxygen vacancies, like in ion-milling \cite{GrossJAP11} or PLD \cite{HerranzAPL09} experiments, has also to be determined. Interestingly enough, a metallic state was recently found when amorphous $LaAlO_3$, $SrTiO_3$ and yttria-stabilized zirconia films are deposited by PLD at room T and under a low oxygen pressure ($<10^{-2}mbar$) on $SrTiO_3$ substrates \cite{ChenNanoLetters11}. This metallic state was also attributed to the presence of oxygen vacancies at the $SrTiO_3$/oxide interface coming from the reduction of the $SrTiO_3$ substrate by the reactive species of the plasma.

\subsection{Comparison with other results}

The room $T$ electronic properties of sample 1 ($R_s$ values, charge carrier density, mobility) are similar to those observed on low charge carrier density $LaAlO_3/SrTiO_3$ heterostructures, when the metallic layer was found to be confined close to the interface \cite{ThielScience06,HuijbenNatMat06}. The differences in the conditions of obtention of the metallic surface between the two systems are however important. First, we don't heat the $SrTiO_3$ substrate whereas a $T$ as high as $800^{\circ}C$ is usually used in PLD experiments during the growth of the $LaAlO_3$ film. Second, no specific preparation of the surface is needed in our case whereas a $TiO_2$ termination was found to be crucial in $LaAlO_3/SrTiO_3$ heterostructures. Last, according to the low $T$ of the substrate during the evaporation, our alumina layer is expected to be amorphous and not crystalline \cite{AmorphousAlumina}. Future structural and physical investigations should clarify the similarities and the differences between the two systems. Ref. \cite{ChenNanoLetters11} has recently demonstrated that a metallic interface can be obtained by the PLD of an amorphous oxide layer at room T. Our results suggest that the use of the PLD technique and a specific surface preparation of the $SrTiO_3$ are even not necessary.

Other results should also be reconsidered in light of our findings. A large but finite resistance was observed in MOSFET devices made by the sputtering of an alumina layer on top of a $SrTiO_3$ surface \cite{UenoAPL03}. It may come from the fact that the deposited alumina layer was slightly oxygen deficient. Similarly, $Al$ is known to make good electrical contacts on the $SrTiO_3$ surface, which may be explained by the existence of a metallic layer below the contacts.

\section{Conclusion}

In conclusion, we have demonstrated that a metallic state can be induced on the $SrTiO_3$ surface by the electron beam evaporation of oxygen deficient alumina or insulating granular aluminium. The resistance of the sample depends on the $Al/O$ ratio during the evaporation process. No heating of the substrate was needed which suggests that the metallic state remains confined close to the surface. Oxygen vacancies seem to be at the origin of this metallicity, which calls for more detailed structural and chemical investigations of the $SrTiO_3$ substrate and its insulating overlayer. We hope that our results and the simplicity of the technique used will stimulate new fundamental and applied research on the electronic properties of the $SrTiO_3$ surface.

This work was financially supported by the R\'egion Rh\^one-Alpes (CIBLE 2010 program) and the Universit\'e Joseph Fourier (SMINGUE 2010 program). Discussions with T. Lopez-Rios are gratefully acknowledged.


\begin{thebibliography}{99}
\bibitem{Contacts} For the two contacts measurements, the samples are channels $250\mu m$ long and $1mm$ wide located between two $Al$ contacts of few $mm^2$. For the four contacts and the Hall effect measurements, a Hall bar geometry was used with arms size and distance between them of $1mm$ and $Al$ contacts of few $mm^2$ connected to the arms.
\bibitem{NoteAl} The heat of formation of $Al_2O_3$ is $520kJ/mol$ $O$ and the work function of $Al$ is about $4.2eV$. It was found that a simple thermodynamic criteria for metal oxidation on $SrTiO_3$ surfaces is a heat of formation (absolute value) larger than $250kJ/mol$ $O$ (see Ref. \cite{FuSSR07}).
\end{thebibliography}
\end{document}